\documentclass[aip,reprint,onecolumn,english,nofootinbib,letterpaper,nobalancelastpage]{revtex4}
\usepackage{lmodern}

\usepackage[T1]{fontenc}
\usepackage[latin9]{inputenc}
\usepackage[letterpaper]{geometry}
\usepackage{amsmath}
\usepackage{amssymb}
\usepackage{bbm}
\usepackage{babel}
\usepackage{graphicx}

\usepackage{latexsym}
\usepackage{mathptm}

\newcommand{\f}{\begin{equation}}
\newcommand{\ff}{\end{equation}}

\begin{document}

\title{{\Large{Gravity's weight on worldline fuzziness}}}
\author{$~$\\
{\bf Giovanni Amelino-Camelia}$^{\, a}$,
  {\bf Valerio Astuti}$^{\, b}$ and {\bf Giacomo Rosati}$^{\, c}$
\\
{\small Dipartimento di Fisica, Universit\`a ``La Sapienza" and Sez.~Roma1 INFN, P.le A. Moro 2, 00185 Roma, Italy } }

\begin{abstract}
{\normalsize \noindent
We investigate a connection between
recent results
in 3D quantum gravity, providing an effective noncommutative-spacetime description,
and some earlier heuristic descriptions of a quantum-gravity contribution to
the fuzziness of the worldlines of particles.
We show that  3D-gravity-inspired
spacetime noncommutativity reflects some of the features
suggested by previous  heuristic arguments.
Most notably, gravity-induced worldline
fuzziness, while irrelevantly small on terrestrial scales,
could be observably large for propagation
of particles over cosmological distances.} \end{abstract}

\maketitle

\vskip 0.3cm

{\large

\baselineskip 14.2pt

Gravitational phenomena weigh on  our daily lives very noticeably,
but are the phenomena whose description
is most unknown at subatomic scales. A fair assessment of the present situation is that we have
access to non-gravitational phenomena down to distance scales of the order of $10^{-20}m$
(LHC scales) whereas we have so far gained
access to gravitational phenomena only at scales no smaller than $10^{-6}m$.
The challenge of quantum-gravity research is accordingly overwhelming:
we have apparently solid indirect evidence (see, {\it e.g.}, Refs.~\cite{carlipREVIEW,gacQM100})
 of the necessity of a new quantum theory of both gravitational and non-gravitational phenomena
with onset at a scale of the order of the minute Planck length  $\ell_P$ ($\sim 10^{-35}m$),
but any experimental guidance we could seek for attempting to describe this new realm of physics
only concerns much larger distance scales.

Over the last decade there has been a determined effort~\cite{gacQM100,gacLRR} attempting
to improve this state of affairs by using the whole Universe as a laboratory.
We focus here on an intriguing example of how this might work out,
in investigations of the ``spacetime-foam" scenario first discussed
 by John Wheeler in the 1960s~\cite{wheeler}
 (also see Refs.~\cite{colemanFOAM,hawkingFOAM,carlipfoam,garayFOAM}).
In some recent studies, such as those in Refs.~\cite{gacgwi,bignapap,ngfoam,ellisfoam,fordBLURRING,clausFOAM},
the spacetime-foam intuition has guided efforts aimed at
characterizing  gravity-induced
contributions to the ``fuzziness" of the worldlines of particles.
One attempts to describe the dynamics of matter particles
as effectively occurring in an ``environment" of short-distance
quantum-gravitational degrees of freedom. And it is expected that
 for propagating particles with wavelength much larger than the Planck length,
when it may be appropriate to integrate out
these quantum-gravitational degrees of freedom,
the main residual effect of short-distance gravity would indeed
be an additional contribution
to the fuzziness of worldlines.
The idea that this might lead to testable predictions originates
from heuristic arguments~\cite{gacgwi,bignapap,ngfoam,ellisfoam,lieuHillman,ragazzoni,ngPRL,tamburini}
suggesting that
these quantum-gravity effects should
grow with propagation distance.
In particular this could produce an observably-large contribution to the
 blurring of the images of distant astrophysical sources,
such as quasars~\cite{ngPRL,tamburini}.

We here do not review the relevant heuristic arguments.
Actually our starting point
is the realization that heuristics was
surely valuable for inspiring this phenomenological program,
but has
run out of steam as a resource for going forward.
This is clear from the ongoing debate
concerning the quantitative assessment of the effects to be sought experimentally.
Essentially this debate revolves around adopting the most promising
phenomenological formula for the description
of the gravity-induced contribution to the uncertainty in the localization
of a particle after propagating over a distance $x$, with two (alternative) such formulas
being considered most actively~\cite{lieuHillman,ragazzoni,ngPRL,tamburini}
\begin{equation}
\delta x\Big|_{grav} \sim \ell_P^\alpha ~ x^{1-\alpha}
\label{fuzzywhite}
\end{equation}
and
\begin{equation}
\delta x\Big|_{grav} \sim \frac{\ell_P^\alpha \,  p^{\alpha}}{\hbar^\alpha} ~ x =
\frac{p^{\alpha}}{M_P^\alpha} ~ x ~.
\label{fuzzyrainbow}
\end{equation}
In these formulas $\ell_P$ denotes again the Planck length, and on the right-hand-side
of (\ref{fuzzyrainbow}) we rendered explicit that $\hbar/\ell_P$ is the ``Planck scale" $M_P$
($\sim 10^{19}GeV$),
$x$ denotes the propagation distance ({\it e.g.}
the distance from a quasar to our observatories), $p$ denotes the momentum
of the particle, and $\alpha$ is a phenomenological parameter, for which the
relevant arguments favor~\cite{lieuHillman,ragazzoni,ngPRL,tamburini}
values between $1/2$ and $1$.

The fact that so far we could only rely on heuristic descriptions
(which also point in rather different directions)
renders it difficult for anyone to form an opinion
on how much effort and resources should be directed
toward developing this phenomenology.
The main objective of this essay is to notice that recent results on quantum gravity
in 3D (2+1-dimensional) spacetime can provide insight on this
from a usefully complementary perspective.
Studies such as the ones reported in Refs.~\cite{kodadsr,kodadsrjureklaurelee,freidelivine,oritiKAPPA}
establish that for 3D quantum gravity (exploiting the much lower complexity than for the 4D case)
we are able to perform the task needed for studies of spacetime fuzziness: we can actually
integrate out gravity, reabsorbing its effects into novel properties
for a gravity-free propagation of particles. And it turns out that this produces a theory
of free particles in a noncommutative spacetime~\cite{kodadsr,kodadsrjureklaurelee,freidelivine,oritiKAPPA},
which in particular could adopt\footnote{While it is established that
the effective spacetime is noncommutative and that it is such that the time coordinate does not commute
with the spatial coordinates, there appears to be still some open issues concerning the proper (or at least
most appropriate) specification of
coordinate noncommutativity~\cite{kodadsr,kodadsrjureklaurelee,freidelivine,oritiKAPPA}.
A noticeable alternative to the $\kappa$-Minkowski coordinates we here adopt is the possibility
of ``spinning coordinates" such that $[x_\mu , x_\nu]= i \epsilon_{\mu \nu \sigma} g^{\sigma \rho} x_\rho$.
The techniques we here develop and use are of general applicability to cases
such that the time coordinate does not commute
with the spatial coordinates, so they could be applied also to studies adopting ``spinning coordinates".}
``$\kappa$-Minkowski" coordinates~\cite{majrue,lukieANNALS}:
\begin{equation}
[x_1 , x_2] = 0~,~~~[x_j , x_0]=i \ell_P x_j~.
\label{kappamink}
\end{equation}
In other words, upon integrating out the gravitational degrees of freedom, the
quantum dynamics of
matter fields coupled to 3D gravity
is effectively described~\cite{kodadsr,kodadsrjureklaurelee,freidelivine,oritiKAPPA}
 by matter fields in
 a noncommutative spacetime.

Our first observation
is that these results on 3D quantum gravity provide some encouragement for the mentioned
hypotheses concerning spacetime fuzziness: at least in the 3D context
one does find that, upon integrating out
the gravitational degrees of freedom, the worldlines of particles acquire
an additional source of fuzziness, since this is surely produced by
 the coordinate noncommutativity.

In work whose preliminary results we here describe, but shall be reported in greater
detail elsewhere~\cite{inprep},
we have exploited this link for characterizing quantitatively
a scenario for gravity's contribution to the fuzziness of worldlines.
For simplicity we focus in this essay on the case of a 2D version of (\ref{kappamink}),
therefore fully characterized by
\begin{equation}
[x_1 , x_0]=i \ell_P x_1 ~.
\label{kappamink2D}
\end{equation}
Our objective is to describe the quantum mechanics of free particles in this spacetime.
And this confronts us immediately with the challenge associated with the fact that
 in $\kappa$-Minkowski the time coordinate is a noncommutative observable,
whereas in the standard formulation of quantum mechanics the time coordinate is merely an evolution
parameter (a necessarily classical evolution parameter).
In the study recently reported in Ref.~\cite{fuzzy1pap}
we advocate the possibility of addressing this challenge by resorting to results obtained
over the last fifteen years
(see, {\it e.g.}, Refs.~\cite{halliwellQM,gambiniportoQM,rovellireisenb})
establishing a covariant formulation of ordinary (first-quantized) quantum mechanics.
In this powerful reformulation of quantum mechanics both the spatial coordinates and the time coordinate
play the same type of role. And there is no ``evolution", since
dynamics is codified in a constraint, just in the same sense
familiar for the covariant formulation of classical mechanics
(see, {\it e.g.}, chapter 4 of Ref.~\cite{henneauxBOOK}).
Spatial and time coordinates are well-defined operators on
a ``kinematical Hilbert space", which is just an ordinary Hilbert space of normalizable
wave functions~\cite{rovellireisenb}. And spatial and time coordinates are still well-defined operators on
the ``physical Hilbert space", obtained from the kinematical Hilbert space
 by enforcing the constraint of vanishing covariant-Hamiltonian.
 Dynamics is codified in the fact that on states of the physical Hilbert space, because
 of the implications of the constraint they satisfy, one finds relationships between the properties
 of the (partial~\cite{rovellireisenb}) observables for spatial coordinates and the properties of
 the time (partial) observable. In this way, for appropriate specification of the state on the physical Hilbert space,
 the covariant pure-constraint version of the quantum mechanics
 of free particles describes ``fuzzy worldlines" (worldlines of particles
 governed by Heisenberg uncertainty principle) just in the same sense that the covariant pure-constraint
 formulation of the classical mechanics of free particles describes sharp-classical worldlines.

This formulation of quantum mechanics is such that
both time and the spatial coordinates are operators on a Hilbert space,
which of course commute among themselves but do not commute with their conjugate momenta,
so that in particular in the 2D case one has~\cite{rovellireisenb}
\begin{gather}
[{\pi}_0,{q}_0] = i \hbar\ , \qquad [{\pi}_0 , {q}_1] = 0 \nonumber \\
[{\pi}_1,{q}_0] = 0 \ , \qquad [{\pi}_1 , {q}_1] = -i \hbar\ ,
\label{pregeomPHSPACE}
\end{gather}

The proposal we put forward in Ref.~\cite{fuzzy1pap}
takes this covariant formulation of quantum mechanics as the starting point for formulating $\kappa$-Minkowski
noncommutativity: the commuting time and spatial-coordinate operators of the covariant formulation
of quantum mechanics should be replaced by
time and spatial-coordinate operators
governed by the $\kappa$-Minkowski noncommutativity.
Specifically we observe in Ref.~\cite{fuzzy1pap} that
the $\kappa$-Minkowski defining commutator (\ref{kappamink2D})
is satisfied
by posing a relationship between $\kappa$-Minkowski coordinates and the phase-space
observables of the covariant formulation of quantum mechanics (the ones of Eq.~(\ref{pregeomPHSPACE}), here viewed
simply as formal auxiliary\footnote{For a different scenario, adopting however
an analogous perspective on role played in the analysis by operators such as $q_0,q_1,\pi_0,\pi_1$,
see Ref.~\cite{meljanacREPRESENTATION}.} operators~\cite{fuzzy1pap})
of the following form:
\begin{eqnarray}
x_1 = \, e^{\ell_P \pi_0/\hbar} \, q_1 ~,~~~x_0 = q_0~.~~~\label{targetX}
\end{eqnarray}
And we also show in Ref.~\cite{fuzzy1pap} that, for consistency with (\ref{targetX}),
one should introduce translation generators $p_1,p_0$
whose action on functions of $\kappa$-Minkowski coordinates, ${x}_0 , {x}_1$,
has the following description in terms of the action of ordinary translation
generators, ${\pi}_{0},\pi_{1}$ on functions of the auxiliary coordinates $q_0 , q_1$:
\begin{gather}
 p_{0}\triangleright f({x}_0 , {x}_1) \longleftrightarrow
  [{\pi}_{0},f({q}_0 ,  {q}_1 e^{\ell {\pi}_0 })] ~,~~~
  p_1 \triangleright f({x}_0 , {x}_1) \longleftrightarrow
   e^{- \ell {\pi}_{0}}[{\pi}_{1},f({q}_0 ,  {q}_1 e^{\ell {\pi}_0 })] ~.
\label{targetP}
\end{gather}
Moreover the ``on-shellness operator" (the operator which, for massless particles,
should vanish on physical states, as enforced
by the Hamiltonian constraint)
should be written in terms of $\pi_1,\pi_0$
of the covariant formulation of quantum mechanics as follows~\cite{fuzzy1pap}
\begin{equation}
 {\cal H}= \left(\frac{2\hbar}{\ell_P}\right)^{2}\sinh^{2}\left(\frac{\ell_P {\pi}_{0}}{2\hbar}\right) - e^{-\ell_P {\pi}_{0}/\hbar} {\pi}_{1}^{2} ~.
\label{dalembertell}
\end{equation}

One more result which is relevant for the observations we are reporting in this manuscript,
among those we established in Ref.~\cite{fuzzy1pap}, concerns the measure for integration
over momenta, needed for evaluating scalar products when working in the ``momentum representation":
we found in Ref.~\cite{fuzzy1pap} that covariance of the $p_0,p_1$-momentum-space integration
 measure
 implies that the $\pi_0,\pi_1$-integration-measure should be $\ell_P$-deformed:
\begin{equation}
 d\pi_0 d\pi_1 \longrightarrow d\pi_0 d\pi_1 e^{-\ell_P \pi_0/\hbar}
\label{measureDEFORMED}
\end{equation}

These results from our previous study Ref.~\cite{fuzzy1pap} were all analyzed there exclusively
on the kinematical Hilbert space. The form of the operator ${\cal H}$ was established,
 but we did not explore the implications of enforcing the Hamiltonian constraint ${\cal H} \Psi_{phys} = 0$
 (for massless particles) in obtaining the physical Hilbert space.
For our purposes here of contributing to the debate on ``gravity's weight on worldline fuzziness"
we must inevitably progress to the next level,  working with the physical Hilbert space,
obtained by enforcing the Hamiltonian constraint.

A key challenge for this objective of the analysis we are here reporting
comes from the fact that the $\kappa$-Minkowski coordinates are not themselves
natural operators for exploring the implications of the physical Hilbert space.
The reason for this indeed comes from the fact that the Hamiltonian constraint
is enforced: the coordinates do not individually\footnote{This challenge is already
present (though in simpler form) in the original commutative-spacetime
setting for the covariant formulation of quantum mechanics. Indeed the Hamiltonian operator $\pi_0^2 - \pi_1^2$
does not commute with $q_0$ and $q_1$ (see Eq.~(\ref{pregeomPHSPACE})).} commute with the Hamiltonian operator.
But this challenge is also an opportunity for the proposal we are here putting forward:
indeed the heuristic arguments supporting one or another {\it ansatz} for ``$\delta x$"
(the ones in Eqs.~(\ref{fuzzywhite}) and (\ref{fuzzyrainbow})) leave some key relativistic issues
unanswered. What does one really mean with the symbol $\delta x$? is that an uncertainty principle
for spatial coordinates? if so, is then the time coordinate immune to this uncertainty principle?

The conceptual perspective of the covariant formulation of quantum mechanics suggests
that uncertainty principles at the most fundamental level are not naturally formulated
as uncertainty principles for single coordinates: again this is due to the fact that a single coordinate
(in our case $x_1$ or $x_0$) does not commute with the Hamiltonian constraint and therefore
is not a ``complete observable"~\cite{rovellireisenb} of the theory.

We propose to remedy this by focusing
on an operator, which we denote by ${\cal A}$, that carries information on the uncertainties
in the spacetime coordinates but does commute with the Hamitonian-constraint operator ${\cal H}$:
\begin{equation}
{\cal A}=e^{\ell_P \pi_0/\hbar} \left( q_1 - {\cal V} q_0 - \frac{1}{2}[q_0,{\cal V}] \right)
\label{calAeq}
\end{equation}
where ${\cal V}$ is short-hand for the operator
$${\cal V} \equiv \left( \frac{\partial {\cal H}}{\partial p^0} \right)^{-1} \frac{\partial {\cal H}}{\partial p^1}$$
which turns out to be such that $e^{\ell_P \pi_0/\hbar}{\cal V}$ plays the role of speed of the particle~\cite{inprep}.

In the classical limit this operator ${\cal A}$ reduces to the observable $x_{1,cl}-v_{cl}x_{0,cl}$
(we place label ``$cl$" on quantities pertaining to the classical limit),
so for the case of free particles we are here considering it gives
the intercept of the particle worldline with the $x_1$ axis.

Because of the special properties of the specific combination of coordinates contained in ${\cal A}$
(particularly the fact that ${\cal A}$ commutes with  the Hamitonian-constraint operator ${\cal H}$)
it is well suited for investigating the issues on which we are here focusing.
Our next task concerns assessing some properties of this observable ${\cal A}$, and
specifically  ``gravity's weight" on $\delta {\cal A}$, {\it i.e.}
the dominant $\ell_P$-induced contribution to $\delta {\cal A}$ which in light of our motivation
is the key objective of this manuscript.
[As announced, we are adopting the working assumption that in the regime here of interest
the effects of quantum-gravitational degrees of freedom are all effectively encoded in
the value of $\ell_P$.]

One other point we need to specify in our formalization of the problem concerns the distance
between source and detector. As emphasized in our opening remarks, the main opportunities provided
by searches of anomalous  blurring of images
of distant quasars should exploit the ``amplifying effect" of the gigantic distance of propagation
from the source (quasar) to our detector (telescope). We 
introduce a dependence on this amplifier by making implicit reference essentially to gaussian
states peaked at ${\bar x}_1,{\bar x}_0$, states which we interpret as describing the case
of a particle emitted from a (fuzzy~\cite{fuzzy1pap}) point
with coordinates ${\bar x}_1,{\bar x}_0$ in the observer's reference frame.
Our first objective is to show that the $\ell_P$-induced contribution to
the uncertainty $\delta {\cal A}$ grows with ${\bar x}_1,{\bar x}_0$,
which will fit with the expectation that $\ell_P$-induced fuzziness 
 grows as the particle propagates over longer and longer distances.

Postponing a more technical analysis~\cite{inprep},
we shall be here satisfied observing that the form of
Eq.~(\ref{calAeq}), keeping in mind in particular that
we are describing $x_1$ as $x_1 = \, e^{\ell_P \pi_0/\hbar} \, q_1$ and $p_0$ as $p_0 =  \pi_0$,
suggests that in the limit of ultralarge ${\bar x}_1$
(and accordingly ultralarge ${\bar x}_0$)
the dominant $\ell_P$-induced contributions must be of order
\begin{equation}
\delta {\cal A}\Big|_{grav}  \sim \frac{\ell_P}{\hbar} {\delta p_0}\, \bar{x}_1
 ~.
\label{sigmatau}
\end{equation}
All other contributions to $\delta {\cal A}\Big|_{grav}$ are either suppressed by higher powers of the
small scale $\ell_P$ or do not benefit from the ``amplification" effectively provided
by the large value of $\bar{x}_1$ (which is indeed very large for the applications
we are here interested in, such as observations of distant quasars).
Eq.~(\ref{sigmatau}) is the main outcome of the analysis we are reporting in this manuscript.
The residual tasks we have concern making contact with the previous heuristic suggestion
for the outcome of such analyses, which we summarized in Eqs.~(\ref{fuzzywhite})-(\ref{fuzzyrainbow}),
and reassessing the outlook of
searches of anomalous  blurring of images
of distant quasars on the basis of this observation.

As stressed above, we feel that our characterization
of spacetime fuzziness through $\delta {\cal A}$
is more powerful than the generic characterization in terms of a ``$\delta x$"
given in formulas such as (\ref{fuzzywhite}) and (\ref{fuzzyrainbow}).
Still we can make some contact between the two characterizations by restricting our focus on cases
of propagation of massless particles such
that $\delta x_1 \gg \delta x_0$ (for some specific observer).
In such cases one concludes from  Eq.~(\ref{calAeq}) that
\begin{equation}
\delta {\cal A}\Big|_{m=0;\delta x_1 \gg \delta x_0}  \sim \delta {x_1}
 ~,
\label{sigmatauDX}
\end{equation}
which we establish also using the fact that for massless particles the uncertainty in ${\cal V}$ vanishes.

In this regime of validity of (\ref{sigmatauDX})
we can rewrite our more general result (\ref{sigmatau})
as follows
 \begin{equation}
\delta x_1\Big|_{grav} \sim \frac{\delta p_0}{M_P} ~ \bar{x}_1
\label{sigmataugravDELTAX}
\end{equation}
 where we replaced $\hbar/\ell_P$ with the Planck scale $M_P$,
 as already done for Eq.~(\ref{fuzzyrainbow}).

Let us incidentally notice that (\ref{sigmataugravDELTAX}) could have been guessed
on the basis of
the noncommutativity relation $[x_1 , x_0]=i \ell_P x_1$, whose form
suggests $\delta x_1 ~\delta x_0 \sim \ell_P \bar{x}_1$; indeed
assuming $\delta x_0 \simeq \hbar / \delta p_0$
(saturating the
Heisenberg uncertainties, as for gaussian states on the Hilbert space)
one obtains from (\ref{sigmataugravDELTAX})
that
 \begin{equation}
\delta x_1\Big|_{grav} \sim \frac{\delta p_0}{M_P} ~ \bar{x}_1
\sim \frac{\ell_P}{\delta x_0} ~ \bar{x}_1
\label{sigmataugravDELTAXbis}
\end{equation}

For what concerns the comparison of our
Eq.~(\ref{sigmataugravDELTAX}) with
the heuristic estimates summarized in Eqs.~(\ref{fuzzywhite}) and (\ref{fuzzyrainbow})
we should start by stressing that none of those parametrized
heuristic estimates of worldline fuzziness corresponds exactly to our result.
But for the mentioned phenomenology, looking for effects blurring the images
of distant quasars~\cite{ngPRL,tamburini,lieuHillman,ragazzoni},
even rough agreement with the estimates
(\ref{fuzzywhite}) or (\ref{fuzzyrainbow})
can be of encouragement. We notice that to the extent that one could argue
for $\delta p_0 \lesssim p_0$
it would be possible to infer from (\ref{sigmataugravDELTAX}) that
 \begin{equation}
\delta x_1\Big|_{grav}
\sim \frac{ \delta p_0}{M_P } ~ \bar{x}_1 \lesssim
\frac{p_0}{M_P} ~ \bar{x}_1 ~.
\label{fuzzyus}
\end{equation}
So there is a rough agreement between our model of spacetime fuzziness and
the heuristic estimate (\ref{fuzzyrainbow})
for the case $\alpha=1$, though our model suggests that
(\ref{fuzzyrainbow}) with $\alpha=1$ should significantly overestimate
the fuzziness (since in general we should
expect $\delta p_0 < p_0$).

In spite of finding only this rough agreement with the case $\alpha=1$
for Eq.~(\ref{fuzzyrainbow}), we feel that we here provided
valuable new tools for attempting to exploit the opportunities
available on the phenomenology side.
The level of fuzziness predicted by  (\ref{fuzzyus}) is truly minute
on terrestrial scales: for example for  a particle with $p \sim 100 GeV$ propagating from preparation to detection over a distance of, say, $x \sim 10^6 m$, testing our
 description of worldline fuzziness would require timing at the detector with the
 unrealistic accuracy
 of $\sim 10^{-20} s$. And yet, as stressed in the opening remarks, these scenarios can be tested
if we use the whole universe as a laboratory.
This is what emerges from the estimates given
in Refs.~\cite{ngPRL,tamburini}
for the associated blurring effects on the images of distant quasars, relying only on
some apparently prudent assumptions concerning the implications of worldline fuzziness
 for an effective randomization of the phase
 of a classical wave (such as the light wave emitted by a quasar).

The work we here reported strengthens the case for
 this phenomenological program, previously describable only through heuristic derivations,
 since we have provided a manageable framework for rigorous derivation
of predictions that can be tested phenomenologically.
The next natural task will be to find ways of
 describing a wave equation within our spacetime-noncommutativity
 setup, so that the link from
  worldline fuzziness to an effective randomization of the phase
 of a classical wave, assumed in  Refs.~\cite{ngPRL,tamburini},
 can also be rigorously scrutinized.

 And of course while the specific type of 3D-gravity-inspired model here adopted
 would favor a scenario somewhat similar to (\ref{fuzzyrainbow})
with $\alpha=1$, we are not implying that other values of $\alpha$
in (\ref{fuzzyrainbow}) or the (\ref{fuzzywhite}) possibility should be disregarded.
On the contrary we believe that, in light of the rare phenomenological window that could be exploited,
all of these pictures should be further investigated. We did however here ``raise the bar" for such studies:
for (\ref{fuzzyrainbow}) with $\alpha \neq 1$ and for the (\ref{fuzzywhite}) case we feel that the most
urgent issue is now finding corresponding manageable quantum-spacetime models,
suitable for taking also the study of those possibilities
at least one step beyond the level of semiheuristic estimates.

}


\end{document}